\documentclass[%
 reprint,
superscriptaddress,
 amsmath,amssymb,
 aps,
prb,
showkeys
]{revtex4-2}

\usepackage{graphicx}
\usepackage{dcolumn}
\usepackage{bm}
\usepackage{multirow}

\usepackage[T1]{fontenc}
\usepackage{graphicx}
\usepackage{siunitx}  
\usepackage[version=4]{mhchem}
\usepackage{breakurl}
\usepackage{libertine}
\usepackage[libertine]{newtxmath}
\usepackage{float}
\usepackage{booktabs}
\usepackage[
colorlinks=true,      
linkcolor=black,       
citecolor=black,       
filecolor=black,       
urlcolor=black,       
pdfpagemode=UseNone  
]{hyperref}
\usepackage{xcolor}
\usepackage{dcolumn}

\begin{document}


\title{Cyclo-Graphyne: A Highly Porous and Semimetallic 2D Carbon Allotrope with Dirac Cones}

\author{Jhionathan de Lima}
\author{Cristiano Francisco Woellner}
\email{woellner@ufpr.br}
\affiliation{Department of Physics, Federal University of Parana, UFPR, Curitiba, PR, 81531-980, Brazil.}
\affiliation{Interdisciplinary Center for Science, Technology, and Innovation (CICTI),  Federal University of Parana,  UFPR, Curitiba, PR, 81530-000, Brazil.}


\begin{abstract}
We present a comprehensive characterization of Cyclo-graphyne (CGY), an emerging 2D carbon allotrope with a porous structure of $\mathrm{sp/sp^2}$-hybridized carbon atoms. Using density functional theory, we systematically investigate its structural, energetic, dynamical, thermal, electronic, mechanical, optical, and vibrational properties. The calculated cohesive and formation energies are both comparable to those of other synthesized graphynes, confirming its energetic viability. Phonon dispersion calculations confirm its dynamical stability, while ab initio molecular dynamics simulations indicate thermal stability up to at least \SI{1000}{\kelvin}. Electronic results reveal that CGY is a semimetal with an ultranarrow band gap and features two Dirac cones in its electronic structure. Mechanically, CGY is highly compliant and isotropic, exhibiting a Young’s modulus an order of magnitude lower than that of graphene. The optical spectrum reveals strong ultraviolet absorption and infrared reflectivity with an isotropic response, while the vibrational spectra show distinct Raman peaks and rich infrared activity. These properties position CGY as a promising candidate for future applications in areas such as gas capture and separation, flexible nanoelectronics, and optoelectronics.
\end{abstract}

\maketitle


\section{Introduction}

Carbon is a remarkably versatile element due to its ability to adopt different hybridizations, including $\mathrm{sp}$, $\mathrm{sp^2}$, and $\mathrm{sp^3}$ configurations. This flexibility enables the formation of a wide variety of carbon allotropes with diverse covalent bonding schemes. Diamond and graphite are the most common natural allotropes of carbon, with hybridization states of $\mathrm{sp^3}$ and $\mathrm{sp^2}$, respectively. Beyond these natural allotropes, a wide range of two-dimensional (2D) synthetic carbon nanomaterials has been theoretically predicted and, in some cases, experimentally realized, including graphene~\cite{Novoselov2004electric}, the biphenylene network~\cite{Fan2021biphenylene}, graphenylene~\cite{Du2017anew}, $\gamma$-graphyne~\cite{Li2018synthesis, Yang2019mechanochemical, Barua2022anovel}, $\gamma$-graphdiyne~\cite{Li2010archtecture}, and the fullerene network~\cite{Hou2022synthesis}.

Among the emerging carbon allotropes, graphynes (GYs) constitute a particularly intriguing class of 2D nanomaterials, as they integrate both $\mathrm{sp}$- and $\mathrm{sp^2}$-hybridized carbon atoms within their extended frameworks. The presence of acetylenic linkages reduces areal density and introduces tunable porosity, leading to distinct physical properties. Several GY allotropes exhibit unconventional electronic band structures characterized by Dirac cones near the Fermi level, including $\alpha$-~\cite{Puigdollers2016first, Hou2018study}, $\beta$-~\cite{Puigdollers2016first, Hou2018study}, $\delta$-~\cite{Zhao2013two}, R-~\cite{Yin2013rgraphyne}, 6,6,12-~\cite{Malko2012competition, Huang2013theexistence}, 14,14,14-~\cite{Huang2013theexistence}, 14,14,18-~\cite{Huang2013theexistence}, 14,12,2-~\cite{Zhang2015highly}, 14,12,4-~\cite{Zhang2015highly}, cp-GY~\cite{Nulakani2017cp}, and $\alpha$-graphdiyne (GDY)~\cite{Niu2013dirac}. For example, the honeycomb $\alpha$-GY exhibits graphene-like dispersion with Dirac cones located at the K and K' points~\cite{Malko2012competition, Huang2013theexistence, Puigdollers2016first, Hou2018study}, although it is energetically less favorable than other variants. In contrast, $\beta$-GY hosts Dirac cones along the $\Gamma$–M path~\cite{Malko2012competition, Huang2013theexistence, Puigdollers2016first}, while 6,6,12-GY presents two nonequivalent, distorted Dirac cones displaced from high-symmetry points, giving rise to intrinsic self-doping effects~\cite{Malko2012competition, Huang2013theexistence}. These unique electronic features highlight GYs as promising candidates for next-generation electronic and optoelectronic applications~\cite{Narang2023areview, Ali2025exploration}.

Motivated by the significant advances achieved through both theoretical predictions and experimental realizations of novel 2D carbon allotropes, considerable effort has been devoted to the search for new members of the GY family with potential for practical applications. Recently, Xu \textit{et al.}~\cite{Xu2022li} investigated the CO$_2$ capture capability of a structure referred to as $\beta 1$-GY. They reported high adsorption capacity when it is decorated with lithium atoms. However, the key fundamental aspects of this structure, such as stability, mechanical response, and electronic structure, remain unaddressed. To the best of our knowledge, no prior work has rigorously examined or characterized this allotrope, leaving a critical gap between its proposed applications and its underlying physical properties.

In this work,  we have conducted the first comprehensive characterization of the carbon allotrope previously reported as $\beta 1$-GY. For clarity and to emphasize its distinctive ring topology, we hereafter refer to this structure as Cyclo-graphyne (CGY). This allotrope is composed of $\mathrm{sp}$- and $\mathrm{sp^2}$-hybridized carbon atoms forming 12-membered rings that alternate between triangular and tetragonal configurations, arranged in a circular motif that gives rise to large 24-membered pores. We systematically investigate its energetic, dynamical, and thermal stability, alongside its electronic, mechanical, optical, and vibrational properties, using first principles calculations based on density functional theory (DFT).

\section{Methodology}

Electronic structure calculations were carried out using the all-electron Fritz Haber Institute Ab-Initio Molecular Simulations (FHI-AIMS)~\cite{Blum2009ab} code. FHI-AIMS expands the electron density and all the operators over an all-electron numerical-atomic-orbitals basis set. In this work, we have used the ``tight'' option, which provides safe pre-constructed default definitions for the different chemical species. The Perdew-Burke-Ernzerhof (PBE)~\cite{Perdew1996generalized} parametrization of the generalized gradient approximation (GGA) of the exchange-correlation energy has been adopted throughout. Recognizing that GGA-PBE significantly underestimates the band gap, a portion of the electronic structure calculations was also done using the hybrid Heyd–Scuseria–Ernzerhof functional (HSE06)~\cite{Heyd2003hybrid}. The Broyden-Fletcher-Goldfarb-Shanno (BFGS) algorithm~\cite{Head1985abroyden} has been used to relax both the atomic coordinates and the lattice vectors, until the maximum force on each atom was below \SI{e{-3}}{\electronvolt\per\angstrom}, and the total energy difference was less than \SI{e{-6}}{\electronvolt}.  A $\Gamma$-centered Monkhorst-Pack k-point mesh of $32\times32\times1$ was used for structural optimization and for calculating electronic, mechanical, optical and vibrational properties. A denser $64\times64\times1$ mesh was employed for the density of states (DOS). A vacuum layer of \SI{20}{\angstrom} along the out-of-plane direction was used to eliminate spurious interactions between periodic images.

AIMD simulations were conducted using the i-PI Python~\cite{Ceriotti2014ipi} code as a universal force and trajectory engine, operating in client–server mode. The forces at each time step were computed on-the-fly using the FHI-AIMS code, and then used by i-PI to update the positions of the nuclei. A Nosé–Hoover thermostat~\cite{Nos1984, Hoover1985} was employed to sample the canonical (NVT) ensemble, with a \SI{1}{\femto\second} integration time step and a total simulation duration of \SI{5}{\pico\second}.

Phonon dispersion relations were obtained using the finite displacement method as implemented in the Phonopy package~\cite{Togo2015first}, with input data derived from DFT simulations. From the fully relaxed unit cell, $2\times2\times1$ supercells incorporating small atomic displacements were generated, and interatomic forces were computed for each configuration. The force constants obtained were then utilized to construct and diagonalize the dynamical matrix, yielding phonon dispersion relations.

To perform optical calculations, the linear macroscopic dielectric tensor $\epsilon_{ij}(\omega)$ was computed within the random phase approximation (RPA) framework~\cite{AmbroschDraxl2006linear}. The imaginary part $\epsilon_2(\omega)$ of the dielectric function was obtained directly from the interband transitions. The corresponding real part $\epsilon_1(\omega)$ was then derived via the Kramers-Kronig transformation~\cite{Waters2000ona}. From these dielectric function components, it is possible to determine key optical properties, such as the absorption coefficient:

\begin{equation}
	\alpha(\omega) = \sqrt{2} \omega\left[ \sqrt{\epsilon_1^2 + \epsilon_2^2} - \epsilon_1 \right]^{1/2},
	\label{eq:absorption}
\end{equation}
the refractive index:
\begin{equation}
	n(\omega) = \dfrac{\sqrt{2}}{2}\left[\sqrt{\epsilon_1^2 + \epsilon_2^2} + \epsilon_1\right]^{1/2},
	\label{eq:refractive}
\end{equation}
and the reﬂectivity:
\begin{equation}
	R(\omega) = \left| \frac{\sqrt{\epsilon_1 + \mathit{i} \epsilon_2} - 1}{\sqrt{\epsilon_1 + \mathit{i} \epsilon_2} + 1} \right|^2.
	\label{eq:reflectance}
\end{equation}
In these equations, $\omega$ is the photon energy.

\section{Results and discussion}
\subsection{Structural properties and energetic stability}

\begin{figure}[h!]
	\centering
	\includegraphics[width=1.0\linewidth]{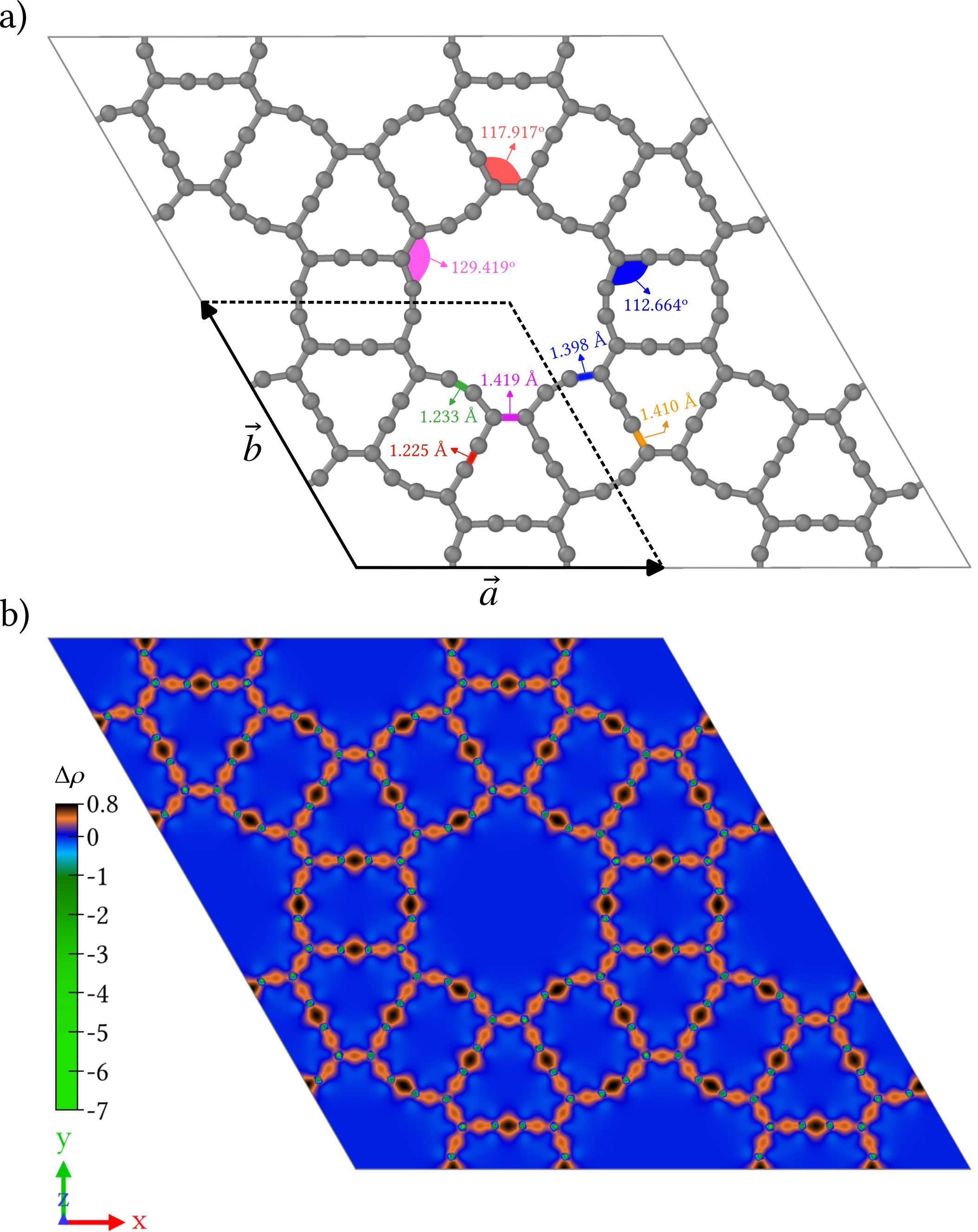}
    \caption{Schematic representation of (a) the CGY monolayer, showing the different \ce{C-C} bond lengths and bond angles. The hexagonal unit cell and lattice vectors are indicated by the black box. (b) charge density difference, highlighting regions of charge accumulation and depletion.}
	\label{fig:structure}
\end{figure}

\autoref{fig:structure}a shows the planar CGY monolayer with a
hexagonal unit cell composed of 36 carbon atoms, belonging to the P6/mmm space group (No. 191). The optimized lattice parameters are $a=b=\SI{13.55}{\angstrom}$, with angles $\alpha=\beta=\SI{90}{\degree}$ and $\gamma=\SI{120}{\degree}$. Unlike graphene, which has a primitive hexagonal lattice with uniform \ce{C-C} bond lengths of \SI{1.422}{\angstrom}~\cite{Trucano1975structure}, the CGY lattice is composed of two distinct types of 12-membered rings that alternate between triangular and tetragonal configurations. Upon periodic repetition of the unit cell, these motifs interconnect in a circular arrangement, giving rise to larger 24-membered pores throughout the lattice. This topology results in five distinct bond lengths ranging from \SI{1.225}{\angstrom} to \SI{1.419}{\angstrom}, and bond angles at the vertices between \SI{112.664}{\degree} and \SI{129.419}{\degree}. The lattice parameters and bond lengths are in good agreement with those reported for $\beta 1$-GY~\cite{Xu2022li}. 

The electronic density distribution was further analyzed through the charge density difference $(\Delta \rho)$, defined as the difference between the self-consistent electronic density and the superposition of the densities of isolated carbon atoms. As illustrated in \autoref{fig:structure}b, positive values are concentrated in interatomic regions, indicative of covalent charge accumulation, with particularly strong localization along the acetylenic linkages. Negative values, localized around the atomic cores, correspond to charge depletion. This pattern highlights the coexistence of both $\mathrm{sp}$ and $\mathrm{sp^2}$ hybridized carbon atoms in the CGY monolayer.

The feasibility of the experimental realization of the CGY nanosheet was evaluated by calculating its formation energy $(E_\mathrm{form})$, defined as:
\begin{equation}
	E_\mathrm{form}=\dfrac{E_\mathrm{total}-N\cdot E_\mathrm{graphene}}{N},
	\label{eq:formation}
\end{equation}
where $E_\mathrm{total}$ is the total energy of the designed CGY monolayer, $E_\mathrm{graphene}$ is the energy per atom in graphene, and $N$ is the total number of carbon atoms in the structure. To evaluate bonding stability, we calculated the cohesive energy $(E_\mathrm{coh})$ using the expression:
\begin{equation}
	E_\mathrm{coh}=\dfrac{E_\mathrm{total}-N\cdot E_\mathrm{C}}{N},
	\label{eq:coesion}
\end{equation}
where $E_\mathrm{C}$ is the energy of an isolated carbon atom. We have calculated the formation and cohesive energies for other 2D hexagnonal carbon allotropes using the same level of theory. In addition, the areal density was computed as the total number of carbon atoms per unit area of the unit cell. The results are presented in \autoref{tab:energies}.

\begin{table}[h]
    \centering
	\caption{Comparison of formation energy ($E_\mathrm{form}$), cohesive energy ($E_\mathrm{coh}$), and areal density among representative 2D carbon allotropes. The values reported in this table were calculated in this study.}

    \label{tab:energies}
    \resizebox{1.0\linewidth}{!}{
        \begin{tabular}{lccc}
            \toprule
            Structure & $E_\mathrm{form}$ (eV/atom) & $E_\mathrm{coh}$ (eV/atom) & Areal density (atom/\SI{}{\angstrom}$^2$)\\
            \midrule
            Graphene      & 0.00 & -9.24 & 0.38\\
			$\gamma$-GY   & 0.64 & -8.60 & 0.29 \\
		    $\gamma$-GDY  & 0.76 & -8.47 & 0.24 \\
            $\beta$-GY    & 0.84 & -8.40 & 0.23 \\
			CGY           & 0.91 & -8.32 & 0.23 \\
			$\alpha$-GY   & 0.92 & -8.31 & 0.19 \\
			Carbyne       & 0.98 & -8.26 & - \\
            \bottomrule
        \end{tabular}
    }
\end{table}

From \autoref{tab:energies}, we conclude that the formation energy of GY-like structures is considerably higher than that of carbon nanomaterials composed solely of $\mathrm{sp^2}$-hybridized carbon atoms, such as graphene. This is attributed to the presence of acetylenic bonds, which are less stable than $\mathrm{sp^2}$ bonds, as evidenced by comparing the formation energies of graphene and carbyne. We further note that the formation energy of CGY is only 0.27 eV/atom and 0.15 eV/atom higher than $\gamma$-GY and $\gamma$-GDY, respectively, both which have already been experimentally synthesized~\cite{Li2018synthesis,Li2010archtecture}. This suggests a positive outlook for the experimental realization of CGY, considering recent advances in synthetic routes for carbon nanomaterials. Additionally, the cohesive energy of CGY is comparable to that of other members of the GY family. These results suggest strong interatomic bonding in CGY, which is essential for maintaining its structural stability. Finally, the low areal densities observed in GY-like structures indicate high porosity, arising directly from the incorporation of acetylenic linkages. Our calculations reveal that CGY possesses a geometric pore size of approximately \SI{8.53}{\angstrom}. This value is about $23\%$ larger than that of $\alpha$-GY (\SI{6.96}{\angstrom}) and nearly three times greater than that of graphene (\SI{2.85}{\angstrom}). Notably, this substantial porosity is achieved without compromising structural integrity, as CGY exhibits a cohesive energy essentially identical to that of $\alpha$-GY. This unique combination of an enlarged pore size and preserved stability positions CGY as a highly promising material for applications in gas capture and separation~\cite{Zhang2012tunable}, energy storage, and water purification~\cite{Kou2014graphyne}.

\subsection{Dynamic and thermal stability}

\begin{figure}[h!]
	\centering
	\includegraphics[width=1.0\linewidth]{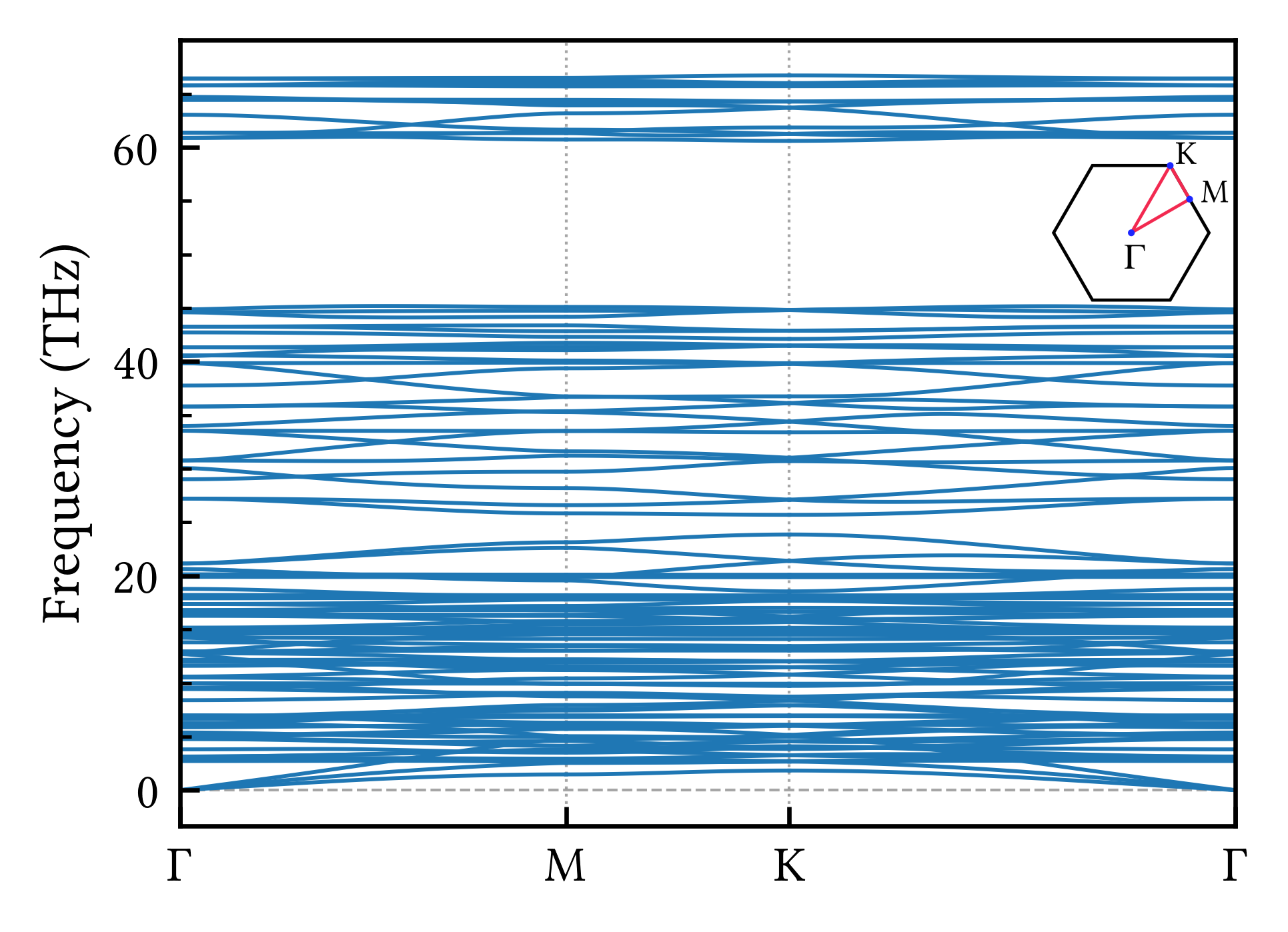}
	\caption{Phonon dispersion relation of CGY along the high-symmetry directions of the Brillouin zone. The absence of negative frequencies across the entire spectrum indicates the dynamical stability of the structure.}
	\label{fig:phonon}
\end{figure}

Phonon calculations are a well-established method for assessing the dynamical stability of theoretical new nanomaterials. \autoref{fig:phonon} presents the phonon dispersion of CGY along high-symmetry points of the Brillouin zone. The absence of phonon branches with negative (imaginary) frequencies across the entire spectrum confirms the dynamical stability of the CGY monolayer. Furthermore, the presence of high-frequency isolated modes in the range from \SI{60}{\tera\hertz} to $\SI{70}{\tera\hertz}$ originates from localized vibrations of the acetylenic bonds. Such modes are commonly observed in other carbon allotropes that contain triple bonds~\cite{Perkgz2014vibrational}. Less dispersive modes appear around \SI{45}{\tera\hertz}, associated with localized vibrations of the remaining carbon atoms in the monolayer. This frequency is comparable to those found in fully $\mathrm{sp^2}$-hybridized carbon nanostructures, such as graphene~\cite{Zou2016phonon}.

\begin{figure}[h!]
	\centering
	\includegraphics[width=1.0\linewidth]{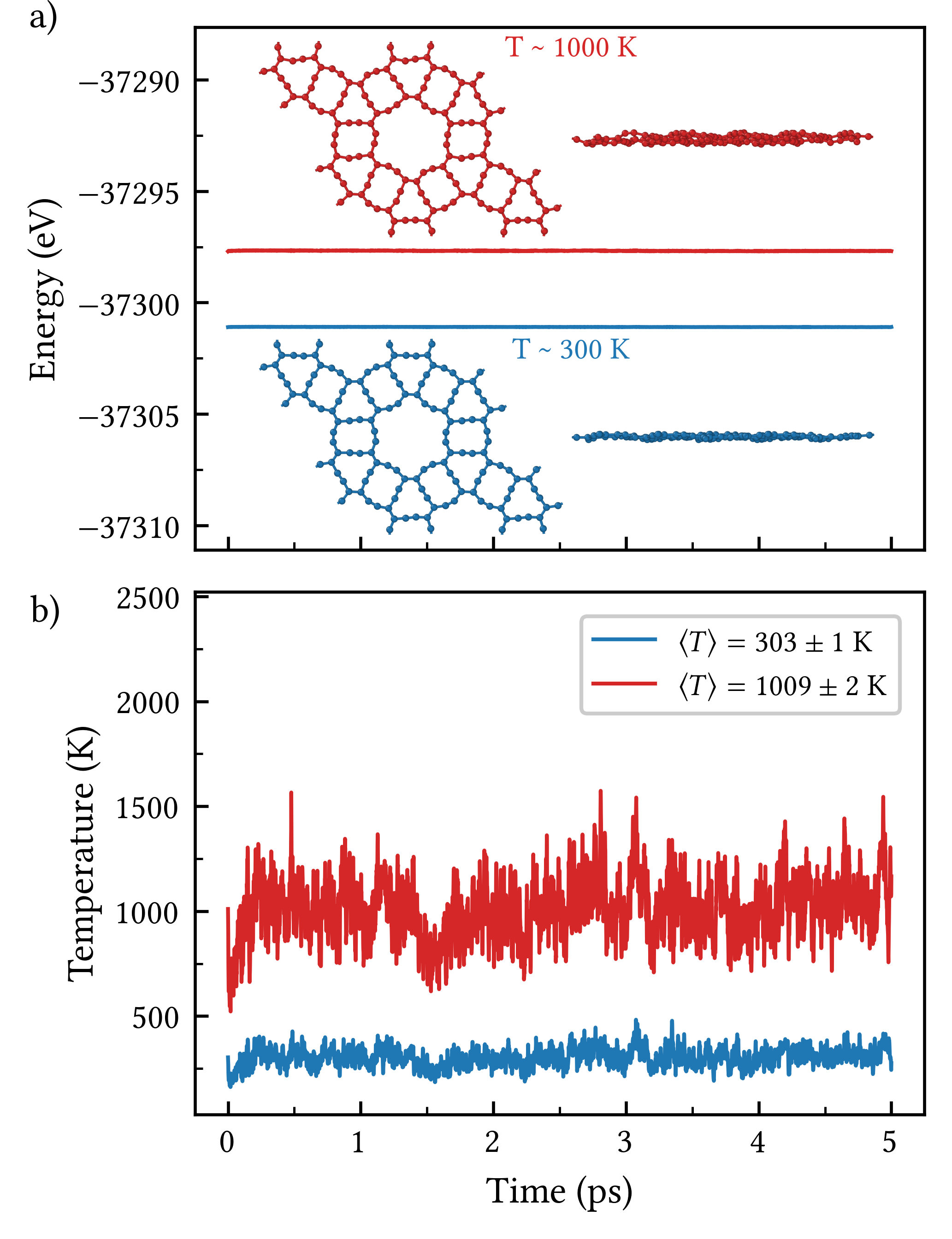}
    \caption{(a) Total energy and (b) temperature fluctuations during AIMD simulations at initial temperatures of \SI{300}{\kelvin} (blue) and \SI{1000}{\kelvin} (red). The insets show the top and side views of the structure at \SI{5}{\pico\second}.}
	\label{fig:aimd}
\end{figure}

The thermal stability of CGY was evaluated through AIMD simulations at finite initial temperatures of \SI{300}{\kelvin} and \SI{1000}{\kelvin}. As illustrated in \autoref{fig:aimd}a, the total energy fluctuates around a steady level for both temperatures, indicating that the system reaches equilibrium without undergoing structural degradation. In addition, the average temperatures shown in \autoref{fig:aimd}b deviate by less than $2\%$ from the target value, reflecting an excellent energy–temperature balance throughout the simulation. These results reveal minimal atomic displacements, no \ce{C-C} bond rupture, and complete preservation of the planar carbon framework (as shown in the insets of \autoref{fig:aimd}a), confirming the thermal stability of the CGY nanosheet up to at least \SI{1000}{\kelvin}.

\subsection{Electronic properties}

\begin{figure}[h!]
	\centering
	\includegraphics[width=1.0\linewidth]{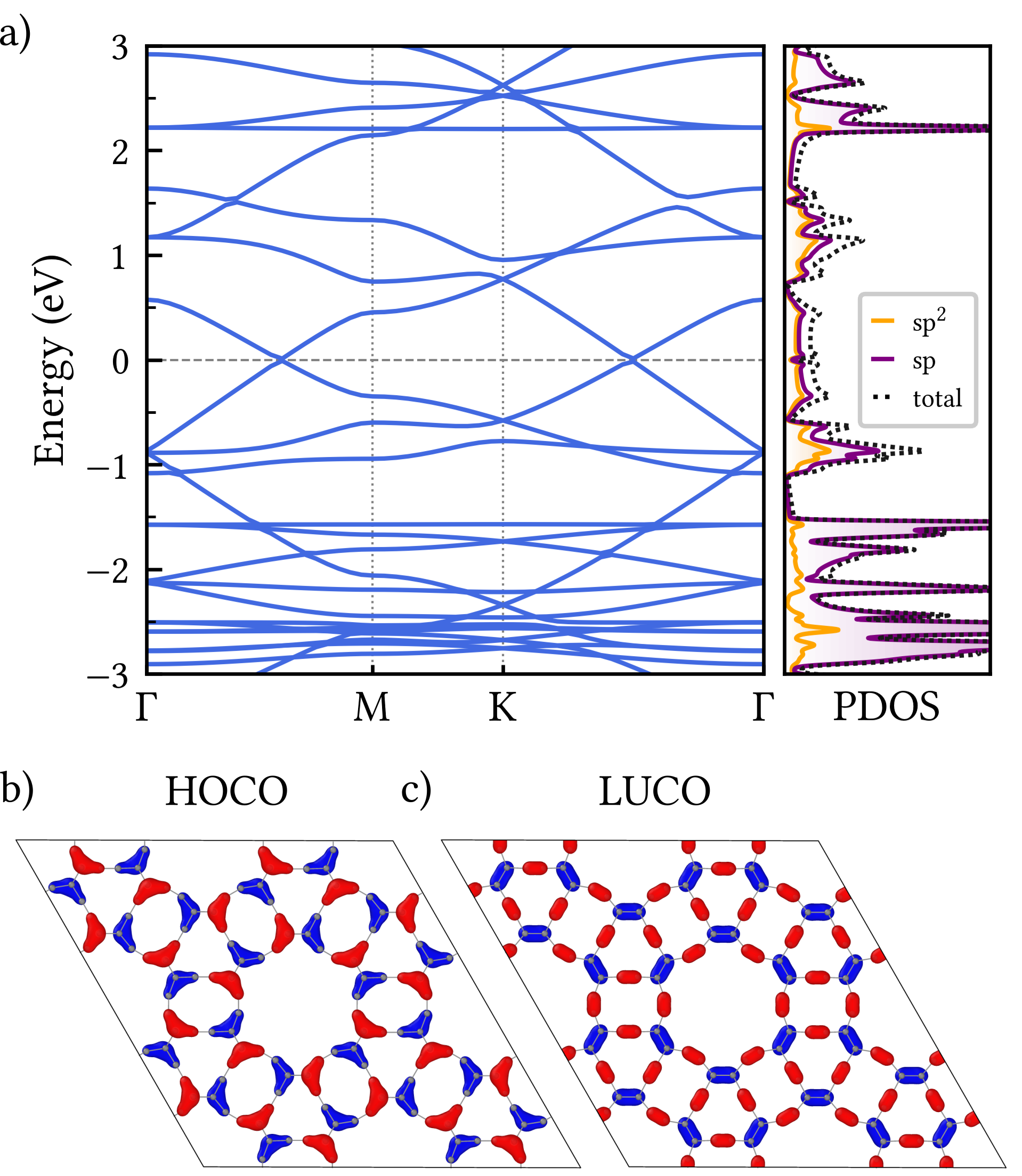}
	\caption{(a) Electronic band structure and projected density of states (PDOS) of the CGY monolayer calculated at the PBE level. The horizontal gray dashed line indicates the Fermi energy. (b–c) Isosurface representations of the highest occupied crystalline orbital (HOCO) and lowest unoccupied crystalline orbital (LUCO), respectively, with red and blue colors denoting different orbital phases.}
	\label{fig:bandpbe}
\end{figure}

The electronic band structure of CGY, calculated at the PBE level, is presented in \autoref{fig:bandpbe}a along the high-symmetry points of the Brillouin zone. The corresponding projected density of states (PDOS) for $\mathrm{sp}$- and $\mathrm{sp^2}$-hybridized carbon atoms, as well as the total density of states, are also shown. A notable feature is the presence of two Dirac points along the $\Gamma\to\text{M}$ and $\text{K}\to\Gamma$ integration paths, indicating that the electrons in CGY behave like massless Dirac fermions, similar to graphene~\cite{Wallace1947theband}. These Dirac points are separated by a narrow gap of approximately \SI{12}{\milli\electronvolt}, revealing the semimetallic nature of the CGY monolayer. The existence of Dirac cones in GYs has been attributed to the effective hopping induced by acetylenic linkages, which renders GY structures topologically equivalent to graphene~\cite{Huang2013theexistence}.

The PDOS analysis from \autoref{fig:bandpbe}a shows that the contributions from $\mathrm{sp}$- and $\mathrm{sp^2}$-hybridized carbon atoms are qualitatively similar within the energy window of \SI{1}{\electronvolt} around the Fermi level. However, the $\mathrm{sp}$ component is naturally larger than the $\mathrm{sp^2}$ one in this region, as we have two $\mathrm{sp}$ carbon atoms for each $\mathrm{sp^2}$ site in the CGY structure.

Visual representations of the frontier electronic states are provided in \autoref{fig:bandpbe}, with panel (b) showing the highest occupied crystalline orbital (HOCO) and panel (c) the lowest unoccupied crystalline orbital (LUCO). These states are the main contributors to reactivity and electronic transport in the vicinity of the Fermi level. It is found that the charge density associated with the HOCO and LUCO are spatially complementary. The HOCO exhibits enhanced amplitude on the $\text{C}^{\mathrm{sp}}-\text{C}^{\mathrm{sp^2}}-\text{C}^{\mathrm{sp}}$ bridges, whereas the LUCO is strongly localized on the $\text{C}^{\mathrm{sp}}\equiv\text{C}^{\mathrm{sp}}$ and $\text{C}^{\mathrm{sp^2}}-\text{C}^{\mathrm{sp^2}}$ bonds. This complementary distribution suggests distinct roles of the acetylenic and mixed hybridized bonds in defining the electronic response of CGY.

\begin{figure}[h!]
	\centering
	\includegraphics[width=1.0\linewidth]{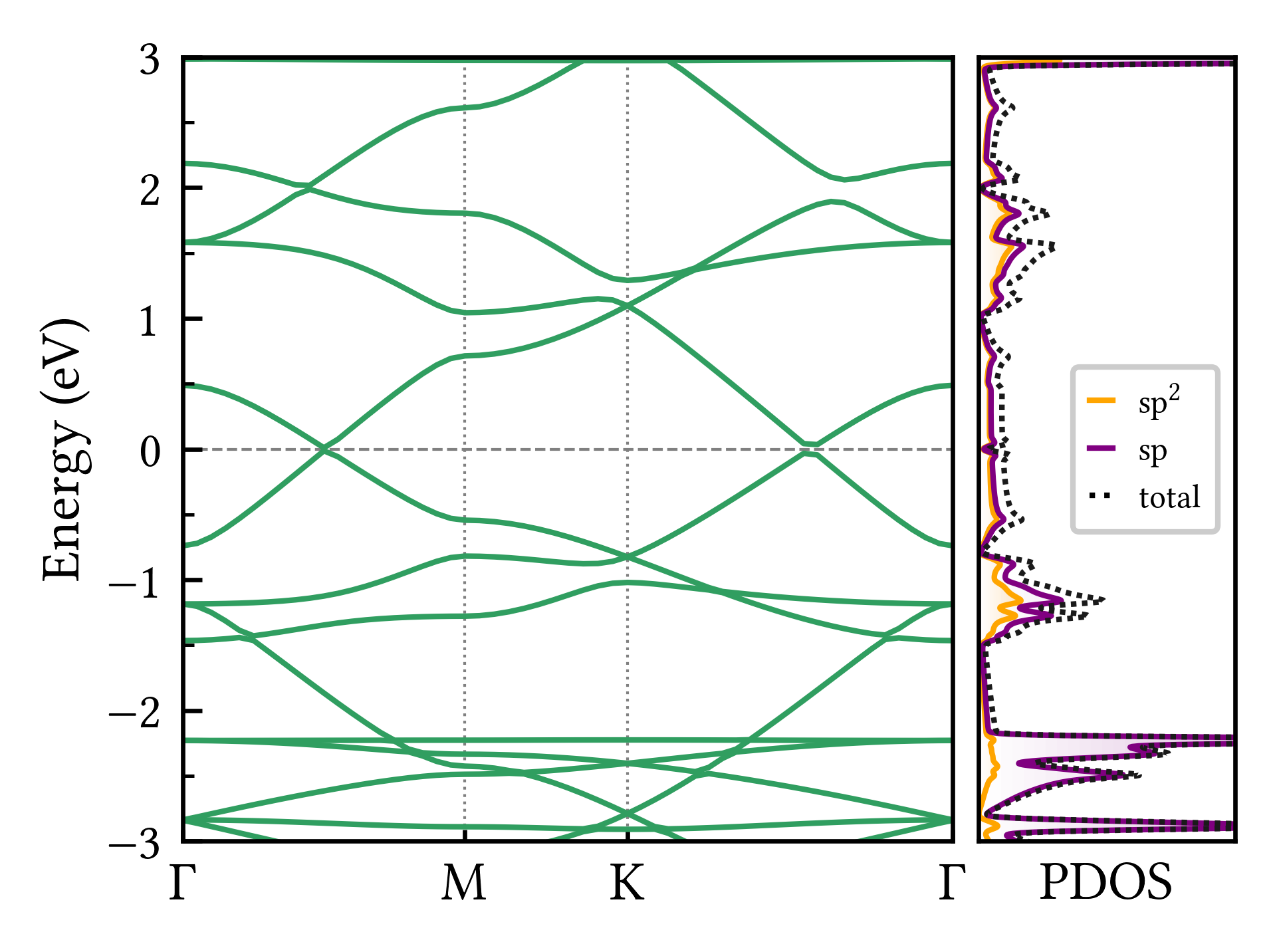}
    \caption{Electronic band structure and projected density of states (PDOS) of the CGY monolayer calculated at the HSE06 level. The horizontal gray dashed line indicates the Fermi energy.}
	\label{fig:bandhse}
\end{figure}

Considering the known limitations of GGA functionals in accurately predicting band gap values, we further computed the electronic band structure of CGY using the HSE06 hybrid functional, as shown in \autoref{fig:bandhse}. The HSE06 results exhibit a band structure qualitatively similar to that obtained from the GGA-PBE calculations (\autoref{fig:bandpbe}a). The most noticeable difference is an almost rigid outward shift of the bands relative to the Fermi energy, with the conduction bands moving upward and the valence bands moving downward. This results in a slightly larger band gap in the HSE06 calculation (\SI{15}{\milli\electronvolt}) compared to the GGA-PBE result (\SI{12}{\milli\electronvolt}). The HSE06 calculations further corroborate the presence of two Dirac cones in CGY, showing no significant gap opening and thereby confirming its semimetallic nature.

\subsection{Mechanical properties}

The mechanical properties of the CGY monolayer were evaluated to assess both its mechanical stability and elastic constants. These calculations were carried out using the energy–strain approach, which analyzes the variation in total energy under small lattice distortions. To this end, in-plane strains were applied by systematically varying the lattice parameters from their unstrained values. We considered strain values in the range $-0.01 \leq \varepsilon \leq 0.01$, with increments of $0.005$. For these small deformations, the elastic strain energy $U(\varepsilon)$, defined as the difference between the total energy of the strained and relaxed systems per unit area, is written in terms of strain components according to the following relation:
\begin{equation}
	U(\varepsilon) = \dfrac{1}{2}C_{11}{\varepsilon_{xx}^2} + \dfrac{1}{2}C_{22}{\varepsilon_{yy}^2} + C_{12}\varepsilon_{xx}\varepsilon_{yy} + 2 C_{66}{\varepsilon_{xy}^2},
	\label{eq:energyxstrain}
\end{equation}
where $C_{11}$, $C_{22}$, $C_{12}$, and $C_{66}$ are the components of the stiffness tensor, corresponding to $1-xx$, $2-yy$, and $6-xy$ according to the Voigt notation~\cite{Andrew2012mechanical}. This general relation is further simplified for isotropic systems with hexagonal symmetry. Such simplification originates from $C_{11}= C_{22}$ and the Cauchy relation $2C_{66}= C_{11}-C_{12}$, yielding~\cite{Cadelano2010elastic}:
\begin{equation}
	U(\varepsilon) = \dfrac{1}{2}C_{11}\left({\varepsilon_{xx}^2 + \varepsilon_{yy}^2 + 2 \varepsilon_{xy}^2}\right) + C_{12}\left({\varepsilon_{xx}\varepsilon_{yy} - \varepsilon_{xy}^2}\right).
	\label{eq:energyxstrainhexagonal}
\end{equation}
From the computed energy–strain curves, $C_{11}$ was extracted via a least-squares fit of the uniaxial deformation data along the $x$ direction. Subsequently, $C_{12}$ was obtained by combining the fitted $C_{11}$ value with the results from equibiaxial strain deformations. For comparison, we also computed the stiffness constants of other 2D hexagonal carbon allotropes at the same level of theory, with the results summarized in \autoref{tab:stiffness_constants}. Unlike bulk materials, 2D systems lack a well-defined thickness, which is required for the calculation of elastic constants in three-dimensional units. Consequently, the elastic constants of 2D structures are conventionally expressed in N/m, rather than in GPa as in bulk materials.

The results in \autoref{tab:stiffness_constants} indicate that the in-plane stiffness constants of the GY-like systems are substantially lower than those of fully $\mathrm{sp^2}$-bonded structures such as graphene. This reduction can be attributed to the presence of acetylenic linkages, which create lower-density frameworks with enhanced structural flexibility. Additionally, as with $\alpha$-GY, the $C_{66}$ value for CGY is significantly smaller than the other stiffness tensor components. This arises from the fact that the unit cell area under shear strain varies within a much smaller range than under uniaxial or biaxial deformations. Combined with the high porosity and structural flexibility of these structures, this explains their very small $C_{66}$ value.

\begin{table}[h]
    \centering
    \caption{Stiffness constants $(C_{ij})$ for selected 2D carbon allotropes. The values reported in this table were calculated in this study.}
    \label{tab:stiffness_constants}
    \resizebox{1.0\linewidth}{!}{
        \begin{tabular}{lcccc}
            \toprule
            Structure & $C_{11}$ (N/m) & $C_{12}$ (N/m) & $C_{66}$ (N/m) & Areal density (atom/\SI{}{\angstrom}$^2$)\\
            \midrule
            Graphene      & 353.26 & 62.59 & 145.33 & 0.38 \\
			$\gamma$-GY   & 201.19 & 84.00 &  58.60 & 0.29 \\
		    $\gamma$-GDY  & 153.81 & 68.60 &  42.61 & 0.24 \\
            $\beta$-GY    & 131.63 & 88.47 &  21.58 & 0.23 \\
			CGY           &  74.42 & 56.94 &   8.74 & 0.23 \\
			$\alpha$-GY   &  95.44 & 83.32 &   6.06 & 0.19 \\
            \bottomrule
        \end{tabular}
    }
\end{table}

The Born–Huang stability criterion~\cite{Mouhat2014necessary} states that, for a 2D material to be mechanically stable, the components of the stiffness tensor must satisfy the conditions $C_{11}C_{22} - C_{12}^{2} > 0$ and $C_{66} > 0$. For isotropic hexagonal systems, these requirements simplify to $C_{11} > |C_{12}|$ and $C_{66} > 0$. As shown in \autoref{tab:stiffness_constants}, the calculated stiffness constants of CGY fulfill these conditions, confirming its mechanical stability.

The components of the stiffness tensor allow for the calculation of fundamental mechanical properties, including the Young’s modulus ($Y$), shear modulus ($G$), and Poisson’s ratio ($\nu$). $Y$ and  $G$ quantify the resistance of the material to uniaxial tensile stress and shear deformation, respectively, within the elastic regime. Furthermore, $\nu$ describes the lateral strain response under uniaxial stress, with lower values indicating less lateral deformation for a given axial strain. Collectively, these parameters reflect the intrinsic bond strength and structural response of the material to mechanical loads. The dependence of $Y$, $G$, and $\nu$ on the $C_{ij}$ components simplifies considerably for isotropic systems~\cite{Polyakova2024elastic}, yielding:

\begin{alignat}{3}
	Y &= \dfrac{C_{11}^2 - C_{12}^2}{C_{11}}, &\quad& G = C_{66}, &\quad& \text{and} \quad \nu = \dfrac{C_{12}}{C_{11}}.
	\label{eq:mechanical_parameters}
\end{alignat}

The calculated elastic constants for CGY are summarized in \autoref{tab:elastic_constants}, alongside values for other 2D carbon allotropes. It is worth noting that the results from \autoref{tab:stiffness_constants} and \autoref{tab:elastic_constants} are in good agreement with those reported in the literature~\cite{Polyakova2024elastic, Ali2025exploration, Puigdollers2016first}. As expected, graphene exhibits the highest Young’s and shear moduli, reflecting its dense, fully $\mathrm{sp^2}$-bonded hexagonal lattice. In contrast, GY- and GDY-based structures display progressively lower $Y$ and $G$ values, which can be attributed to the incorporation of acetylenic linkages.

Further analysis of the data in \autoref{tab:elastic_constants} indicates that CGY exhibits intermediate mechanical properties within the GY family, with a Young’s modulus of $Y = \SI{30.85}{\newton\per\meter}$, a shear modulus of $G = \SI{8.74}{\newton\per\meter}$, and a relatively high Poisson’s ratio of $\nu = 0.77$. The reduction in $Y$ and $G$, compared to graphene, reflects the effect of its porous structure and extended acetylenic chains, which allow the lattice to deform more readily under applied stress. The large $\nu$ indicates a pronounced lateral response to uniaxial stress, further highlighting its flexibility. As illustrated in \autoref{fig:structure}a, the $-\text{C} \equiv \text{C}-$ bridges in CGY are slightly curved, particularly near the tetragonal pores. This geometry suggests that under biaxial strain, deformations are accommodated primarily through bond-angle variations rather than direct bond stretching or compression. Consequently, the $C_{11}$ and $C_{12}$ components of the stiffness tensor are very close in value, resulting in a high $\nu$.

\begin{table}[h]
    \centering
    \caption{The values of Young’s modulus $(Y)$, shear modulus $(G)$, and Poisson’s ratio $(\nu)$ for selected 2D carbon allotropes. The values reported in this table were calculated in this study.}
    \label{tab:elastic_constants}
    \resizebox{1.0\linewidth}{!}{
        \begin{tabular}{lrrcc}
            \toprule
            Structure & $Y$ (N/m) & $G$ (N/m) & $\nu$ & Areal density (atom/\SI{}{\angstrom}$^2$)\\
            \midrule
            Graphene      & 342.17 & 145.33 & 0.18 & 0.38 \\
			$\gamma$-GY   & 166.12 &  58.60 & 0.42 & 0.29 \\
		    $\gamma$-GDY  & 123.21 &  42.61 & 0.45 & 0.24 \\
            $\beta$-GY    &  72.17 &  21.58 & 0.67 & 0.23 \\
			CGY           &  30.85 &   8.74 & 0.77 & 0.23 \\
			$\alpha$-GY   &  22.70 &   6.06 & 0.87 & 0.19 \\
            \bottomrule
        \end{tabular}
    }
\end{table}

\begin{figure}[h!]
	\centering
	\includegraphics[width=1.0\linewidth]{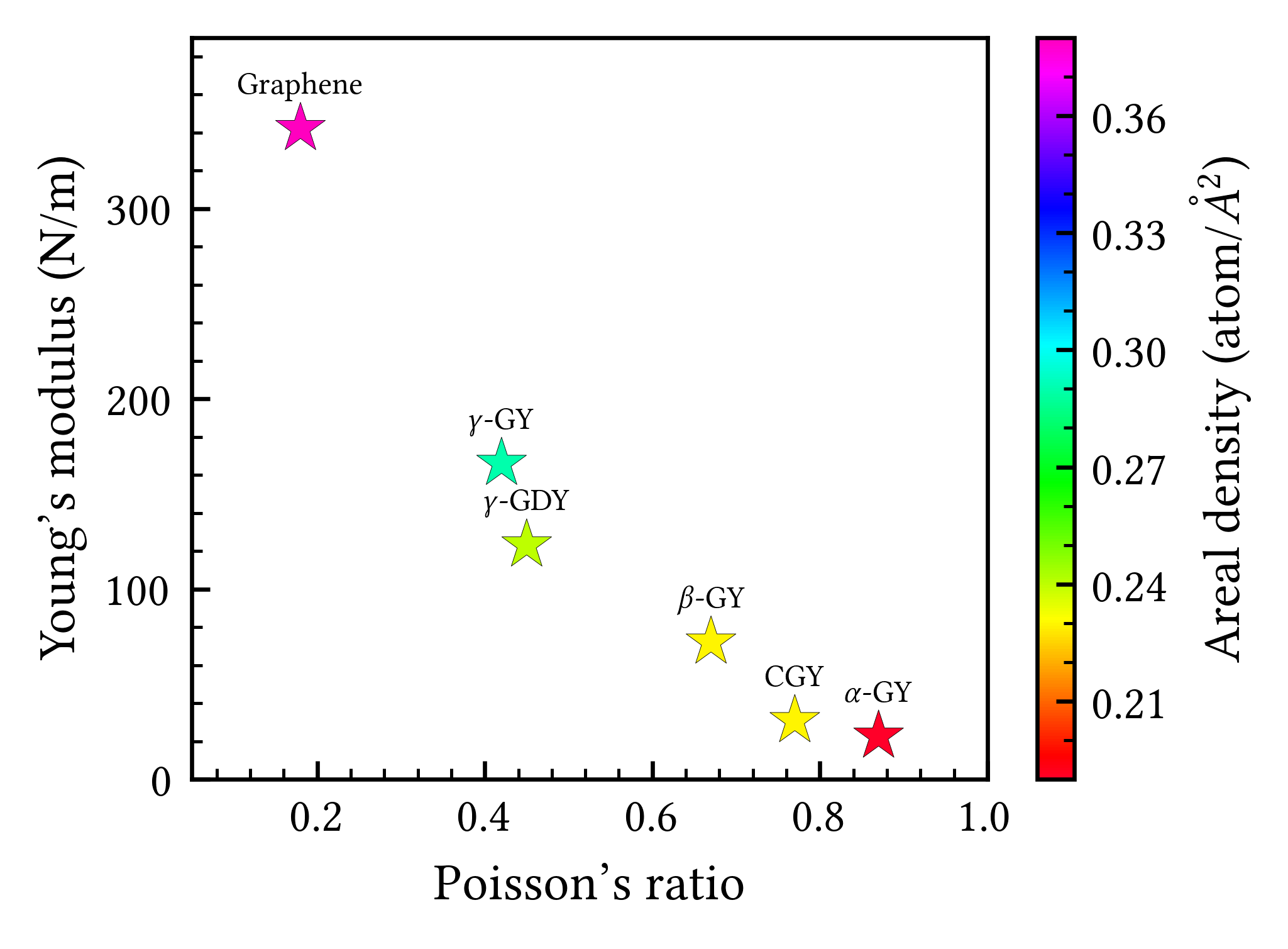}
	\caption{Young’s modulus versus Poisson’s ratio for selected 2D carbon allotropes. The colorbar represents the areal density of each structure.}
	\label{fig:mechanical}
\end{figure}

The trend relating elastic constants to areal density, as observed in \autoref{tab:elastic_constants}, is further illustrated in \autoref{fig:mechanical}. The plot of Young’s modulus versus Poisson’s ratio, colored according to areal density, reveals a continuous trend across the GY family, where decreasing areal density is associated with reduced stiffness and enhanced flexibility. The incorporation of acetylenic linkages introduces porosity, reduces the areal density, and enhances structural flexibility, thereby decreasing the in-plane elastic constants. These findings confirm that the mechanical response of GY monolayers is governed by the presence and spatial arrangement of acetylenic linkages, which distinguishes them from fully $\mathrm{sp^2}$-bonded carbon allotropes such as graphene.

\subsection{Optical properties}

\begin{figure}[h!]
	\centering
	\includegraphics[width=1.0\linewidth]{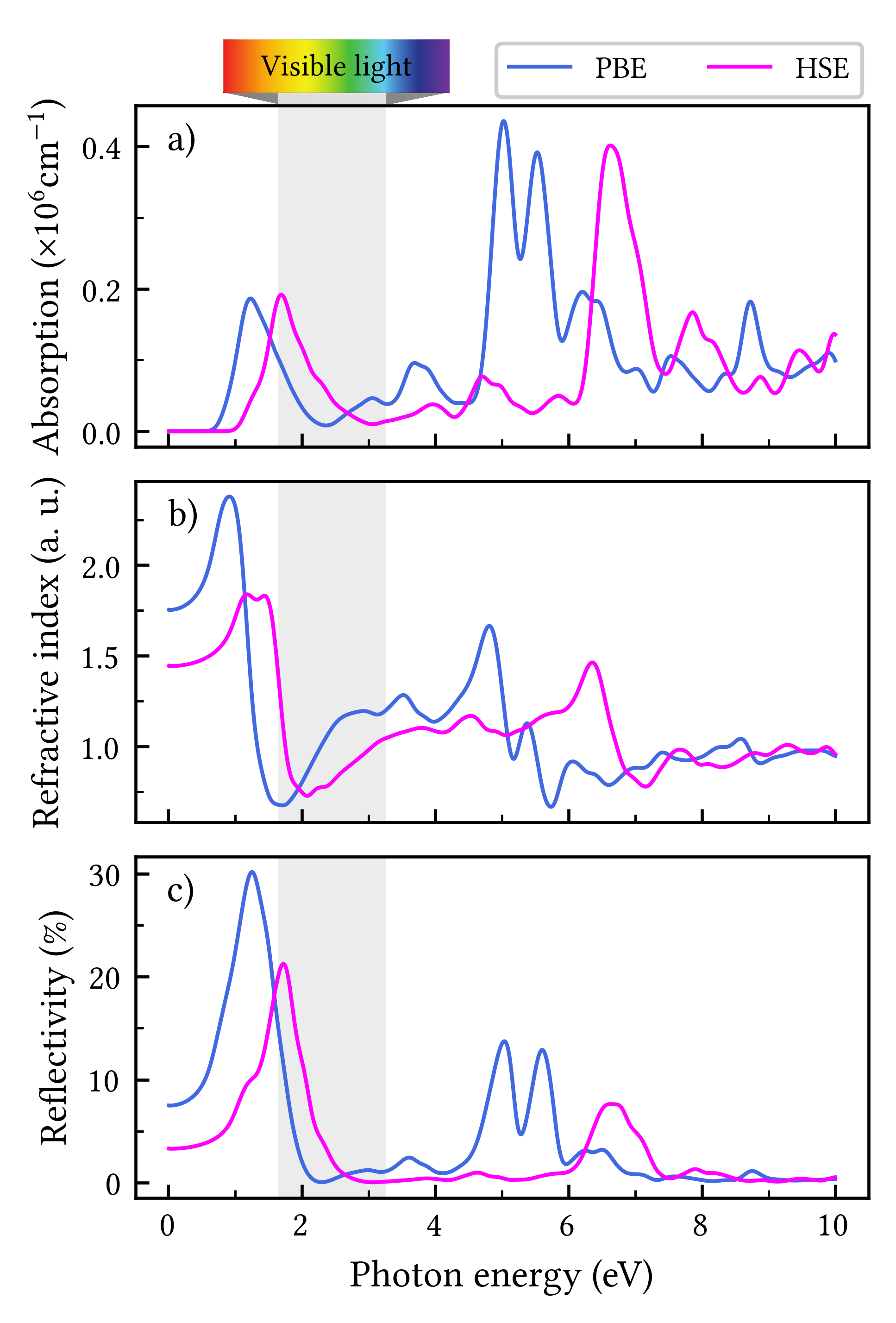}
    \caption{In-plane optical properties of the CGY monolayer: (a) absorption coefficient, (b) refractive index, and (c) reflectivity as a function of photon energy.}
	\label{fig:optical}
\end{figure}

In \autoref{fig:optical}, the optical coefficients of CGY are presented as a function of photon energy. Due to the nearly isotropic nature of the monolayer, the in-plane optical response along the $x$ and $y$ directions is essentially identical. Unlike other GY allotropes, where absorption spectrum displays finite intensity already near zero photon energy~\cite{Hou2018study}, CGY exhibits negligible absorption below \SI{0.7}{\electronvolt} (\autoref{fig:optical}a). The first prominent feature corresponds to a sharp peak in the near-infrared (IR) region, followed by strong absorption in the ultraviolet (UV). This optical profile suggests that CGY could be particularly suitable for applications in photodetectors and optical sensors.

The refractive index (\autoref{fig:optical}b) reaches its highest values in the IR region, with lower values in the UV and visible ranges. Similarly, the reflectivity (\autoref{fig:optical}c) is highest in the IR region, reaching approximately \SI{30}{\percent}. It drops sharply beyond \SI{1.5}{\electronvolt} before increasing again between \SI{4}{\electronvolt} and \SI{7}{\electronvolt}. The overall magnitude of the CGY reflectivity is moderate compared to the high reflectivity of $\beta$- and $\gamma$-GY, and it is instead similar to that of $\alpha$-GY~\cite{Hou2018study}.

For all optical coefficients decipeted in \autoref{fig:optical}, the HSE06 spectrum is rigidly shifted by approximately \SI{0.5}{\electronvolt} to higher energies relative to PBE. This shift is consistent with the known tendency of hybrid functionals to increase the band gap and partially correct the underestimated transition energies of semilocal functionals.

\subsection{Vibrational properties}

\begin{figure}[h!]
	\centering
	\includegraphics[width=1.0\linewidth]{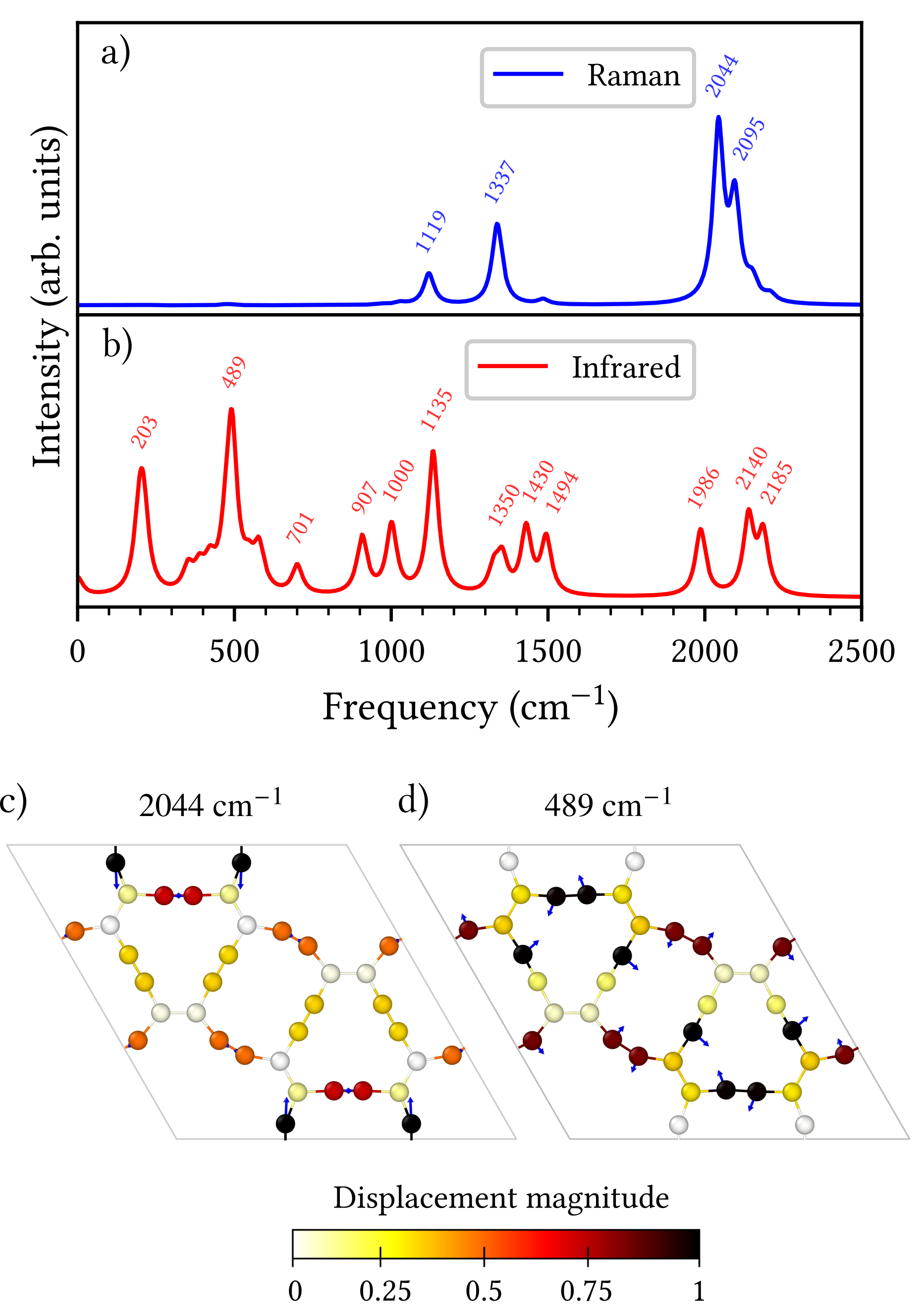}
    \caption{Simulated (a) Raman and (b) infrared spectra of the CGY monolayer with labeled peak frequencies (in \SI{}{\centi\meter^{-1}}). Panels (c) and (d) show the atomic displacement patterns corresponding to the most intense Raman and infrared peaks, respectively. Normalized displacement magnitudes are represented by a white-to-black color scale (inactive to active), and blue arrows indicate the displacement directions.} 
	\label{fig:ramanandif}
\end{figure}

Raman and infrared (IR) spectroscopies are powerful experimental techniques for probing bonding characteristics and lattice dynamics in carbon-based materials. Accordingly, the calculated vibrational spectra for CGY, shown in \autoref{fig:ramanandif}, provide a direct fingerprint for future experimental characterization. The Raman spectrum presented in \autoref{fig:ramanandif}a is characterized by four prominent peaks located at \SI{1119}{\centi\meter^{-1}}, \SI{1337}{\centi\meter^{-1}}, \SI{2044}{\centi\meter^{-1}}, and \SI{2095}{\centi\meter^{-1}}. The first two low-frequency modes are related to symmetric in-plane vibrations of $\mathrm{sp^2}$-hybridized carbon atoms. In contrast, the higher-frequency peaks are unambiguous signatures of \ce{C#C} stretching vibrations, a characteristic trait of GYs structures~\cite{Popov2013theoretical}. The atomic displacement pattern for the most intense mode at \SI{2044}{\centi\meter^{-1}}, visualized in panel (c), confirms its assignment to symmetric stretching along the triple bonds.

The IR spectrum shown in \autoref{fig:ramanandif}b exhibits a richer distribution of vibrational modes compared to the Raman spectrum. The lowest-frequency peak at \SI{203}{\centi\meter^{-1}} is attributed to out-of-plane vibrations. The most intense band in the entire spectrum, located at \SI{489}{\centi\meter^{-1}}, along with a feature at \SI{701}{\centi\meter^{-1}}, arise from symmetric and asymmetric bending vibrations involving the acetylenic linkages. The atomic displacement pattern for this intense \SI{489}{\centi\meter^{-1}} mode is shown in panel (d), confirming its character. Numerous bands between \SI{907}{\centi\meter^{-1}} and \SI{1494}{\centi\meter^{-1}} are assigned to collective bending and stretching motions of both $\mathrm{sp}$- and $\mathrm{sp^2}$-hybridized carbon atoms. Finally, strong absorption features in the high-frequency region at \SI{1986}{\centi\meter^{-1}}, \SI{2140}{\centi\meter^{-1}}, and \SI{2185}{\centi\meter^{-1}} are unequivocally assigned to \ce{C#C} bond stretching vibrations.

Taken together, the calculated Raman and IR spectra from \autoref{fig:ramanandif} provide a comprehensive vibrational fingerprint unique to the CGY monolayer. The distinct high-frequency \ce{C#C} stretching modes, particularly the strong Raman peak at \SI{2044}{\centi\meter^{-1}}, and the intense IR peak at \SI{489}{\centi\meter^{-1}}, serve as unambiguous spectroscopic signatures for identifying this allotrope. These results not only elucidate the lattice dynamics and bonding character of CGY but also establish a crucial reference for its future experimental characterization and differentiation from other GY variants.

\section{Summary and conclusions}

In this work, we have provided a comprehensive characterization of Cyclo-graphyne (CGY), an emerging two-dimensional carbon allotrope composed of $\mathrm{sp}$- and $\mathrm{sp^2}$-hybridized carbon atoms. Its structure forms 12-membered rings that alternate between triangular and tetragonal configurations, arranged in a circular motif that gives rise to large 24-membered pores. First-principles all-electron calculations confirm its energetic stability, while phonon dispersion curves and AIMD simulations demonstrate its dynamical and thermal stability up to at least \SI{1000}{\kelvin}.

Electronic structure calculations reveal that CGY is a semimetal with a narrow band gap, featuring two Dirac cones. Its hexagonal lattice incorporating acetylenic linkages gives rise to nearly isotropic elastic behavior. The high porosity originating from its acetylenic linkages results in a highly compliant structure, with a Young’s modulus approximately 11 times lower and a Poisson’s ratio nearly 4 times higher than those of graphene. Optical simulations reveal a nearly isotropic in-plane response, characterized by strong ultraviolet absorption and pronounced infrared reflectivity. Furthermore, the calculated vibrational spectra provide a unique fingerprint for experimental identification, featuring a few sharp, distinct Raman peaks complemented by a richer array of infrared modes.

In summary, our results highlight CGY as a mechanically flexible yet structurally robust 2D carbon allotrope. Its unique structure, featuring large 24-membered pores, positions CGY as a promising candidate for future applications in areas such as gas capture and separation, flexible nanoelectronics, and optoelectronics.

\begin{acknowledgments}
We thank the Coaraci Supercomputer for computer time (Fapesp grant \#2019/17874-0) and the Center for Computing in Engineering and Sciences at Unicamp (Fapesp grant \#2013/08293-7). We also acknowledge the National Laboratory for Scientific Computing  (LNCC/MCTI, Brazil) for providing HPC resources of the SDumont supercomputer, which have contributed to the research results presented in this work. C.F.W acknowledges financial support from Coordination for the Improvement of Higher Education Personnel (CAPES), the Brazilian National Council for Scientific and Technological Development (CNPq), and the Araucaria Foundation.
\end{acknowledgments}



\bibliography{manuscript.bib}
\end{document}